\documentclass[12pt]{article}
\usepackage{amssymb}
\usepackage{graphicx}
\usepackage{makeidx}
\usepackage[dvips]{epsfig}
\usepackage{multicol}

\def\Journal#1#2#3#4{{#1} {\bf #2}, #3 (#4)}
\def \PRL      {Phys. Rev. Lett.~}
\def \PR       {Phys. Rev.}

\def \PLB      {Phys. Lett. B}
\def \ZPC      {Z. Phys. C}     

\def \PR       {Phys. Rep.~}

\def \CPC      {Comput. Phys. Commun.}
\def \EUR      {Eur. Phys. J. C}
\def \etal     {\relax\ifmmode{et \; al.}\else{$et \; al.$}\fi}

\begin{document}

\title{GR@PPA\_4b: A Four Bottom Quark Production Event Generator 
	for $pp$/$p\bar{p}$ Collisions}

\author{S. Tsuno\footnote{Corresponding author, e-mail : tsuno@fnal.gov} \ and 
	K. Sato \\
        {\itshape Institute of Physics, University of Tsukuba} \\
        {\itshape Tsukuba, Ibaraki 305-8571, Japan} \\
        {} \\
        J. Fujimoto, T. Ishikawa, Y. Kurihara, S. Odaka and Y. Takaiwa \\
        {\itshape High Energy Accelerator Research Organization(KEK)} \\
        {\itshape Tsukuba, Ibaraki 305-0801, Japan} \\
        {} \\
        T. Abe \\
        {\itshape Graduate School/Faculty of Science, Tohoku University} \\
        {\itshape Miyagi 980-8578, Japan} \\
        }


\maketitle

\vspace*{-0.8cm}
\begin{abstract}
	We have developed an event generator, named GR@PPA\_4b, for the four 
	bottom quark ($b\bar{b}b\bar{b}$) production processes at $pp$ and 
	$p\bar{p}$ collisions. The program implements all of the possible 
	processes at the tree level within the framework of the Standard 
	Model. Users can generate events from the Higgs boson and $\gamma/Z$ 
	mediated processes, as well as those from pure QCD interactions. The 
	integration and the event generation are performed within the newly 
	developed GR@PPA framework, an extension of the GRACE automatic 
	event-generator generation system to hadron collisions. This program 
	is so designed that it can be embedded in a general-purpose event 
	generator PYTHIA version 6.1. PYTHIA adds the initial- and final-state 
	parton showers and simulates the hadronization and decays to make 
	generated events realistic. It should be emphasized that a huge number 
	of diagrams and complicated four-body kinematics are dealt with 
	strictly in GR@PPA\_4b. This program will be useful for studies of 
	Higgs boson productions, especially those in extended models where the 
	Yukawa coupling to $b$ quarks is greatly enhanced.

	The source code is located in 
	{\tt http://atlas.kek.jp/physics/nlo-wg/index.html}.

\end{abstract}

\vspace*{-23cm}
\begin{flushright}
KEK Preprint 2002-7 \\
KEK-CP-122 $\qquad\quad\;$ \\
UTPP-68 $\qquad\qquad\quad$ \\
\end{flushright}

\clearpage


\leftline{{\large\bf PROGRAM SUMMARY}}

\vspace*{1cm}
\hspace*{-0.8cm}
\begin{minipage}[t]{.47\textwidth}
\textit{Title of program:} GR@PPA\_4b (v1.0) \\

\textit{Program obtainable from:} {\tt http://atlas} \\
{\tt .kek.jp/physics/nlo-wg/index.html} \\

\textit{Operating system under which the program has been tested:} UNIX \\

\textit{Programming language used:} Fortran77 \\

\textit{Memory required to execute with typical data:} 56.5 kwords for an 
integration, 74.6 kwords for an event generation \\

\textit{Number of bytes in distributed program, including test data, etc.:} 
3153920 \\

\textit{Distribution format:} tar gzip file \\

\textit{Keywords:} GR@PPA, GRACE, PYTHIA, Higgs, bottom quark, $pp$/$p\bar{p}$ 
collisions \\

\textit{Nature of physical problem} \\
Four bottom-quark production is an important channel for the study of 
Higgs-boson properties at future high energy hadron-collider experiments. 
However, the detectability is very ambiguous because only crude estimates 
based on many \\
\end{minipage}\hfill\begin{minipage}[t]{.47\textwidth}
approximations have been available for 
background processes. \\


\textit{Method of solution} \\
GR@PPA\_4b calculates the cross section and generates unweighted events of 
four bottom-quark production in $pp$/$p\bar{p}$ collisions, on the basis of 
exact matrix elements of all the possible processes at the tree level within 
the Standard Model. QCD processes as well as the Higgs boson and $\gamma$/$Z$ 
mediated processes are included. The program has been developed within the 
framework of GR@PPA, an extension of the GRACE system to hadron collisions, 
and embedded in PYTHIA. \\


\textit{Restrictions on the complexity of the program} \\
The Yukawa couplings of lighter quarks ($u$, $d$, $s$ and $c$) are ignored. 
The bottom-quark content in the beam hadrons is not taken into account. \\


\textit{Typical running time} \\
2 hours for a cross-section integration and 200 msec per 1 event for an event 
generation.
\end{minipage}

\section{Introduction}

	Despite the remarkable success of the Standard Model (SM) in high 
	energy physics during the recent decades, nothing is known about the 
	source of its fundamental theoretical basis, the Higgs mechanism, 
	because the Higgs bosons, the remnants of this mechanism, remain 
	undiscovered. The search for the Higgs bosons is thus considered to be 
	the most important subject in the Tevatron Run II \cite{tevhiggs} and 
	forthcoming LHC \cite{lhchiggs} experiments.

	The properties of the Higgs bosons, thus promising search channels, 
	depend on the underlying theory. The minimal supersymmetric extension 
	of the Standard Model (MSSM) \cite{mssm}, which is considered to be a 
	promising theory to solve difficulties in the SM, requires the 
	existence of three neutral Higgs bosons. Among them, the CP-odd one 
	and, in many cases, one of the two CP-even ones, have appreciably 
	large couplings to the bottom quark over a wide parameter range (large 
	$\tan\beta$ regions). The production associated with a bottom quark 
	pair is a promising search channel in this case.

	These Higgs bosons with large couplings to the bottom quark 
	predominantly decay to a bottom quark pair. Therefore, this process 
	can be experimentally tagged as four bottom-quark events, and the 
	Higgs boson production can be identified by a resonant enhancement in 
	the invariant mass spectrum of two bottom quarks. In spite of such a 
	clear signature, a discovery in this way is not trivial because of 
	the presence of the huge QCD background \cite{dai,valls}. Actually, a 
	previous study for LHC \cite{richter-was} showed a discouraging 
	result. Because only crude estimates based on many approximations have 
	been available for the background processes, the prospects are still 
	quite ambiguous.

	In order to provide more reliable tools for this kind of studies, we 
	have developed a Monte Carlo event generator of four bottom-quark 
	productions from $pp$ and $p\bar{p}$ collisions. The program, named 
	GR@PPA\_4b, calculates the cross section and generates realistic 
	(unweighted) events, based on a complete tree-level calculation of 
	all possible processes within the Standard Model, including QCD 
	processes as well as the Higgs boson and $\gamma$/$Z$ mediated 
	processes. The results can be applied to MSSM cases by changing the 
	normalization for the Higgs boson-mediated processes according to the 
	change of the coupling strength.

	The core part of the program, describing parton-level hard 
	interactions, was generated by using an automatic calculation system, 
	called GRACE \cite{grace}. Because the GRACE system has been developed 
	mainly aiming at applications to lepton collisions, generated codes 
	are not directly applicable to hadron-collision interactions. We have 
	developed an extended framework, called GR@PPA (GRACE at PP/Anti-p), 
	to implement those features specific to hadron collisions 
	\cite{odaka}. The primary function of GR@PPA is to determine the 
	initial-state partons, $i.e.$ their flavors and momenta, by referring 
	to a parton distribution function (PDF). Since the GR@PPA framework is 
	not process-specific, it can be applied to any other processes in 
	hadron collisions.

	Based on the GRACE output codes, GR@PPA calculates the cross section 
	and generates unweighted parton-level events using 
	BASES/SPRING \cite{bases} included in the GRACE system. The GR@PPA 
	framework also includes an interface to a general-purpose event 
	generator, PYTHIA version 6.1 \cite{pythia}. Using this 
	interface\footnote{A similar extension of GRACE has been realized in a 
	previous work, GRAPE \cite{abe} for $ep$ collisions. In the present 
	work we adopt a different method (an embedding method), expecting an 
	improvement in the usability of the program.}, the GR@PPA program can 
	be totally embedded in a PYTHIA program. The generated parton-level 
	event information, including the color flow, is automatically passed 
	to PYTHIA. The initial- and final-state radiation, hadronization and 
	decays can be implemented by PYTHIA, to make the generated events 
	realistic.

	This paper is organized as follows: the GR@PPA extension of the GRACE 
	system is described in Section 2. The features of GR@PPA\_4b are 
	specified in Section 3. All details about running the program are 
	given there. Some physical results and program performances are 
	presented in Section 4. A summary is given in Section 5. Typical 
	Feynman diagrams of the processes implemented in the program are 
	shown in the appendix.

\section{GR@PPA}

\subsection{Extension of GRACE to $pp$/$p\bar{p}$ collisions}

	Cross sections with a hard interaction in $pp$/$p\bar{p}$ collisions 
	can be described as

\begin{equation}
	\sigma = \sum_{i, j, F} \int dx_{1} \int dx_{2} \int d\hat{\Phi}_{F} 
	f^{1}_{i}(x_{1},Q^{2}) f^{2}_{j}(x_{2},Q^{2}) 
	{ d\hat{\sigma}_{i j \rightarrow F}(\hat{s}) \over d\hat{\Phi}_{F} },
	\label{eq:xsec}
\end{equation}

	where $f^{a}_{i}(x_{a},Q^{2})$ is a PDF of the hadron 
	$a$ ($p$ or $\bar{p}$), which gives the probability to find the parton 
	$i$ with an energy fraction $x_{a}$ at a probing virtuality of 
	$Q^{2}$. The differential cross section 
	$d\hat{\sigma}_{i j \rightarrow F}(\hat{s})/d\hat{\Phi}_{F}$ describes 
	the parton-level hard interaction producing the final-state $F$ from a 
	collision of partons, $i$ and $j$, where $\hat{s}$ is the square of 
	the total initial 4-momentum. The sum is taken over all relevant 
	combinations of $i$, $j$ and $F$. Note that in hadron interactions a 
	certain "process" of interest may contain some incoherent subprocesses 
	having different final states, as well as those having different 
	combinations of the initial-state partons. For example, the "two-jet" 
	production process includes all $q\bar{q'}$, $qg$($\bar{q}g$) and $gg$ 
	production processes.

	The original GRACE system assumes that both the initial and final 
	states are well-defined. Hence, it can be applied to evaluating 
	$d\hat{\sigma}_{i j \rightarrow F}(\hat{s})/d\hat{\Phi}_{F}$ and its 
	integration over the final-state phase space $\hat{\Phi}_{F}$ only. 
	An adequate extension is necessary to take into account the variation 
	of the initial state both in parton species and their momenta, in 
	order to make the GRACE system applicable to hadron collisions.

\begin{figure}[htbp]
\begin{center}
\includegraphics[width=14.0cm]{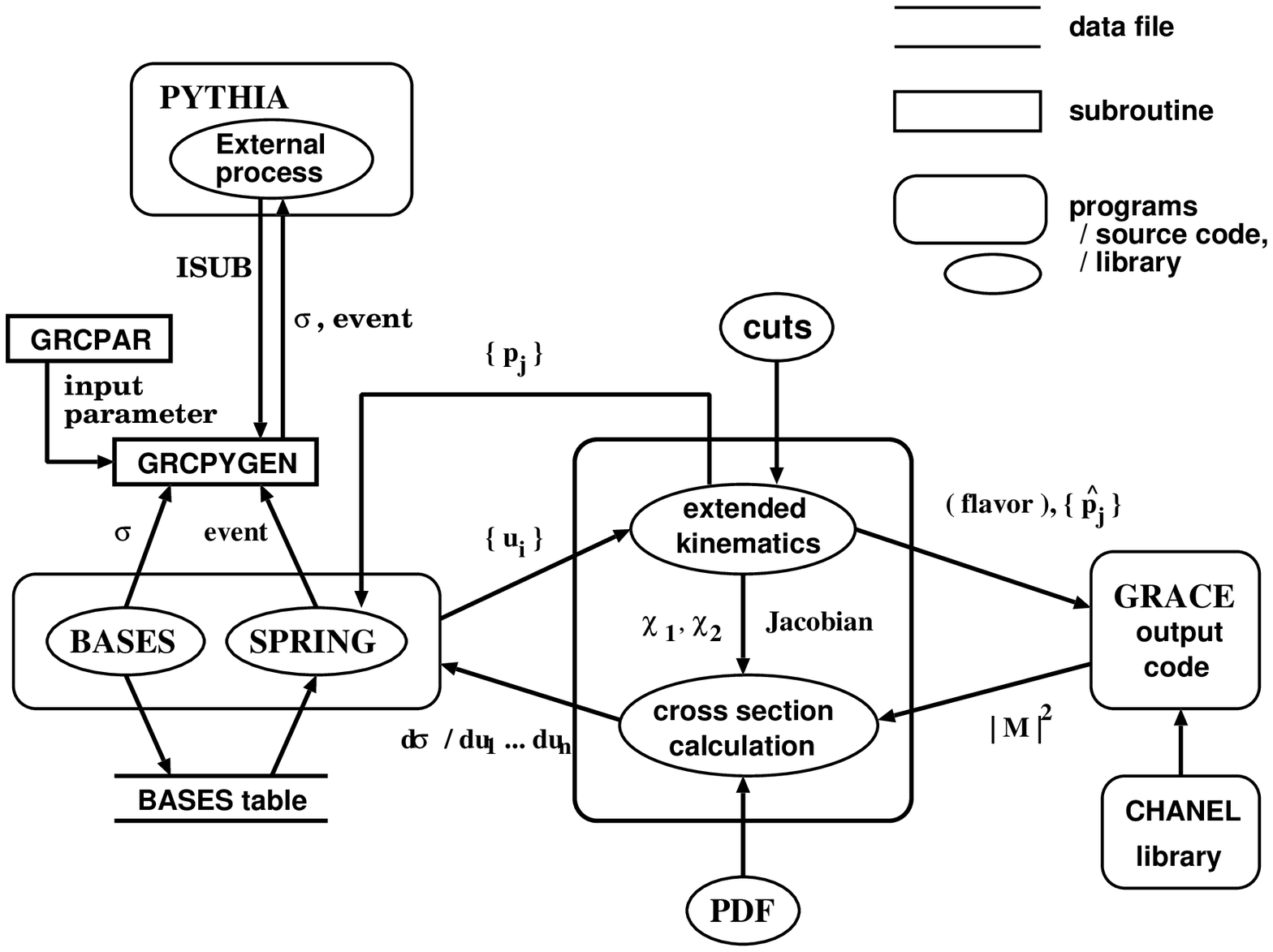}
\caption{Picture showing the structure of GR@PPA. The data flow is indicated 
	schematically. The main building blocks of the GRACE-based event 
	generator are BASES/SPRING and "GRACE output codes". The interface 
	between them has been extended for $pp$ and $p\bar{p}$ collisions. A 
	set of random numbers \{$u_{i}$\} given by BASES/SPRING includes two 
	numbers for defining the initial state ($x_{1}$ and $x_{2}$), in 
	addition to those for defining the final state. The cross section is 
	calculated from the matrix element returned from the "GRACE output 
	codes", by referring to a PDF using $x_{1}$ and $x_{2}$. Some 
	phase-space cuts are applied by limiting the range of kinematical 
	variables, or by setting the cross section to zero after the 
	kinematical variables are determined. In some cases, several 
	subprocesses are combined to a single subprocess by adding one more 
	random number for defining the quark flavor. This system is interfaced 
	to PYTHIA 6.1 through {\tt grcpygen} and {\tt PYUPEV}.}
\label{fig:grappa}
\end{center}
\end{figure}

	The structure of the GR@PPA system is schematically drawn in 
	Fig.~\ref{fig:grappa}. The basic elements of the system, which are the 
	same as the original GRACE system, are the "GRACE output code" and 
	BASES/SPRING. The "GRACE output code" is a set of FORTRAN 
	codes for calculating the matrix element of a specified process, 
	according to a set of kinematical variables specifying a phase-space 
	point. The codes can be automatically generated using a utility 
	included in the GRACE system. The codes for four-$b$ production 
	processes have been generated by the authors and included in the 
	GR@PPA\_4b distribution. 

	BASES/SPRING is a multi-dimensional general-purpose Monte Carlo 
	integration and event-generation program set. It generates a set of 
	random numbers to give them to an external function. Using the 
	returned answer, BASES performs an integration and SPRING generates 
	"events" by means of a hit/miss method. The most remarkable feature of 
	the BASES/SPRING system is the utilization of a multi-dimensional grid 
	method for the random number generation. BASES optimizes the grid 
	setting by an iteration to maximize the efficiency of the integration 
	and the event generation. The optimized setting is stored in an 
	external file (BASES table) to be used later in the event generation 
	by SPRING.

	The remaining task required to GRACE users is to prepare the interface 
	between BASES/SPRING and the "GRACE output code". The interface has to 
	convert the random numbers given by BASES/SPRING to a set of 
	kinematical variables necessary for the matrix element calculation 
	("kinematics"), and to convert the returned matrix element to the 
	differential cross section. Singular structures such as the $1/k$ 
	singularity of the photon/gluon radiation and Breit-Wigner resonance 
	structures, has to be taken into account in the conversion to the 
	kinematical variables, using their well-known asymptotic forms. 
	Although the grid method of BASES/SPRING is very flexible and 
	practically very powerful, by itself, it is not capable of dealing 
	with these singularities without any care.

	An extension has been made in the interface between BASES/SPRING and 
	the GRACE output code. We require BASES/SPRING to provide two 
	additional random numbers, in order to determine the initial-state 
	variables, $x_{1}$ and $x_{2}$. Due to a $1/x$ asymptotic behavior of 
	the structure functions, it is convenient for Monte Carlo integration 
	and event generation to rewrite Eq.~(\ref{eq:xsec}) as

\begin{equation}
	\sigma = \sum_{i, j, F} \int \frac{d\tau}{\tau} \int dy 
	\int d\hat{\Phi}_{F} 
	x_{1}f^{1}_{i}(x_{1},Q^{2}) x_{2}f^{2}_{j}(x_{2},Q^{2}) 
	{ d\hat{\sigma}_{i j \rightarrow F}(\hat{s}) \over d\hat{\Phi}_{F} },
	\label{eq:xsec1}
\end{equation}

	where

\begin{equation}
	\tau = x_{1} x_{2} \; , \quad \quad 
	y = \frac{1}{2} \ln \frac{x_{1}}{x_{2}}.
	\label{eq:tau-y}
\end{equation}

	In GR@PPA, the added two random numbers are converted to $\tau$ and 
	$y$, while taking into account the asymptotic $1/\tau$ behavior for 
	$\tau$, and assuming a flat probability distribution for $y$. The 
	variable $\tau$ determines the center-of-mass (cm) energy of the 
	hard interaction, since $\hat{s}$ $=$ $\tau s$. The variables $x_{1}$ 
	and $x_{2}$ are derived using Eq.~(\ref{eq:tau-y}) in order to refer 
	to PDFs in the conversion of the returned matrix element to the 
	differential cross section, as shown in Fig.~\ref{fig:grappa}. The 
	interface finally returns the calculated differential cross section to 
	BASES/SPRING, and at the same time converts the kinematical variables 
	in the cm frame to ones in the laboratory frame, by applying a 
	Lorentz boost determined by $y$.

	As already mentioned, a "process" of interest is usually composed of 
	several incoherent subprocesses in hadron interactions. However, the 
	present version of BASES/SPRING can treat only one subprocess at the 
	same time. This does not matter in BASES. It is sufficient to do the 
	integration and the grid optimization sequentially for these 
	subprocesses, one after the other. On the other hand, this is a 
	serious limitation in event generation by SPRING, because we 
	frequently want to generate events of different subprocesses in a 
	random order.

	We applied a slight modification to SPRING to overcome this 
	difficulty. The "BASES table" is prepared for every subprocess by 
	running BASES sequentially over the subprocesses. The modified SPRING 
	works as follows: when SPRING is called at the first time, all 
	relevant "BASES tables" are read into a tentative memory area. The 
	main "BASES table" to be used for random-number generation is replaced 
	in each event, by copying an appropriate one from the tentative 
	memory. This method works well because entire information specific to 
	subprocesses, such as the optimized grid information and the cross 
	section information, is recorded in the "BASES table".

	Although we successfully extended the BASES/SPRING to multiple 
	subprocesses, the number of subprocesses is desired to be as small 
	as possible because we have to prepare not only the "BASES table", but 
	also the "GRACE output codes" for every subprocess. In many cases, 
	the difference between the subprocesses is the difference in the 
	quark combination in the initial and/or final states only. The matrix 
	element of these subprocesses is frequently identical, or the 
	difference is only in a few coupling parameters and/or masses. In such 
	cases, it is convenient to add one more integration/differentiation 
	variable to replace the summation in Eqs.~(\ref{eq:xsec}) and 
	(\ref{eq:xsec1}) with an integration. As a result, these subprocesses 
	can share an identical "GRACE output code" and can be treated as a 
	single subprocess by BASES/SPRING. This extension is implemented in 
	GR@PPA\_4b for $q\bar{q} \rightarrow b\bar{b}b\bar{b}$ subprocesses.

\subsection{Interface to PYTHIA}

	As shown in Fig.~\ref{fig:grappa}, GR@PPA includes an interface to a 
	general-purpose event generator PYTHIA version 6.1. Using a facility 
	in PYTHIA, we can add the effect of initial- and final-state parton 
	showers to the generated events. This effect emerges as a finite 
	overall $p_{T}$ of the hard interaction system and finite underlying 
	activities. Furthermore, if we activate the hadronization and decay, 
	we can obtain realistic events which can be passed to detector 
	simulators. 

	We use the subroutine {\tt PYUPEV}, prepared by PYTHIA to deal with 
	external generators, as the interface in the PYTHIA side. The prepared 
	{\tt PYUPEV} simply calls the GR@PPA steering routine {\tt grcpygen}. 
	The subroutine {\tt grcpygen} controls BASES/SPRING and, as a result, 
	controls all GR@PPA routines.

	When {\tt PYUPEV} is used to generate events of user-defined 
	processes, PYTHIA requires users to specify the estimated maximum 
	cross section {\tt SIGMAX} for each process in the initialization 
	stage by using the subroutine {\tt PYUPIN}. {\tt PYUPEV} is required 
	to return the normalized cross section {\tt SIGEV} for each event. The 
	{\tt SIGEV} is so defined that the average should become the total 
	cross section. The ratio {\tt SIGEV}/{\tt SIGMAX} is the weight of 
	this event. PYTHIA determines "accept or reject" of the event using 
	this weight.

	The subroutine {\tt grcpygen} calls BASES or SPRING according to the 
	mode selection determined by an input argument. In GR@PPA, users must 
	call {\tt grcpygen} in the initialization stage before calling the 
	PYTHIA initialization by {\tt PYINIT}, with the mode selection for 
	calling BASES to evaluate the total cross section. In this call, 
	{\tt grcpygen} internally calls {\tt PYUPIN} by setting the 
	argument {\tt SIGMAX} equal to the evaluated total cross section. In 
	the event generation cycle, {\tt PYUPEV} calls {\tt grcpygen} with the 
	mode selection for calling SPRING. Since the event generation is 
	totally controlled by SPRING in GR@PPA, the rejection in PYTHIA must 
	be deactivated. For this purpose, the returned argument of 
	{\tt grcpygen}, which is directly passed to the argument {\tt SIGEV} 
	of {\tt PYUPEV}, is always set equal to the total cross section 
	evaluated by BASES.

	The calling sequence of {\tt grcpygen} is as follows: 

	{\tt call grcpygen(beams, ISUB, mode, sigma)}, 

	where the input arguments are

\begin{tabbing}
	{\tt beams (CHARACTER)} \= : '{\tt PP}' for $pp$ collisions 
	and '{\tt PAP}' for $p\bar{p}$ collisions \\
	{\tt ISUB (INTEGER)} \> : subprocess number \\
	{\tt mode (INTEGER)} 
	\> : = 1 for calling BASES, and 0 for calling SPRING,
\end{tabbing}

	and the output is

	{\tt sigma (REAL*8)} : integrated cross section.

	The argument {\tt beams} is a dummy when {\tt mode} = 0. PYTHIA 
	requires users to assign a unique subprocess number {\tt ISUB} to 
	every user-defined subprocess. The output {\tt sigma} is always equal 
	to the integrated cross section of the subprocess specified by 
	{\tt ISUB}.

	The most important task of {\tt grcpygen} in the event generation 
	cycle is to pass the event information determined in GR@PPA to PYTHIA. 
	The interfacing rules are all specified by PYTHIA. The information 
	concerning the parton species and momenta, which has been determined 
	in the "kinematics" routines and passed through the user interface 
	routine of SPRING, is copied to the arrays in the common {\tt PYUPPR}. 
	The color flow information, which is necessary to perform 
	hadronization, is also recorded, based on the information from 
	SPRING \cite{grc4f}.

\section{GR@PPA\_4b}

\subsection{Subprocesses}

	Based on the GR@PPA framework, we have developed an event generator, 
	called GR@PPA\_4b, for the production of four bottom quarks 
	($b\bar{b}b\bar{b}$) in $pp$ and $p\bar{p}$ collisions. The 
	calculations are all done within the framework of the minimal 
	Standard Model. We divide the process into eight subprocesses 
	according to the difference in the initial state and the order of the 
	couplings, as listed in Table~\ref{tab:subprocess}. The subprocesses 
	are listed in the order of the subprocess number ({\tt ISUB}). We 
	assigned those numbers reserved in PYTHIA for user-defined processes. 
	The number of included Feynman diagrams is also listed in the table 
	for each subprocess. We can see that a large number of diagrams which 
	are hard to manage manually are included.

	In GR@PPA\_4b we do not account for bottom quarks in the initial 
	state; namely, only the lighter quark ($u$, $d$, $s$ and $c$) pairs, 
	as well as the gluon pairs, are counted as the initial state. Since 
	the lighter quarks do not appear in the final state, the functional 
	form of the matrix element is identical for all quark-initiated 
	subprocesses. We treat these subprocesses as a single subprocess, by 
	adding one variable for the choice of the initial-state quark flavor, 
	as explained in a previous section. We generated the "GRACE output 
	code" for each of these combined subprocesses. Note that the 
	interference between those diagrams belonging to different 
	subprocesses are ignored in GR@PPA\_4b.

	We include all order-four tree level interactions within the Standard 
	Model in this generator. Typical diagrams are shown in Appendix A. The 
	symbol $y_{b}$ in Table~\ref{tab:subprocess} represents the Yukawa 
	coupling between the Higgs boson and the bottom quark. We ignore the 
	Yukawa couplings of lighter quarks. Those subprocesses including 
	$y_{b}^{2}$ are composed of diagrams including a bottom-quark pair 
	production mediated by the Higgs boson. Namely, they are the "signal" 
	processes according to our present interest.

	The strong and electroweak couplings are symbolically represented by 
	$\alpha_{s}$ and $\alpha_{em}$, respectively. The subprocesses 
	classified as $\alpha_{s}^{2}\alpha_{em}^{2}$ include irreducible 
	"$Z^{0}$ background". Those classified as $\alpha_{s}^{4}$ are the 
	non-resonant but most serious "QCD backgrounds". The contribution of 
	the subprocesses classified as $\alpha_{em}^{2}y_{b}^{2}$ and 
	$\alpha_{em}^{4}$ is expected to be small but included for the 
	completeness.

\subsection{Distribution package}

	The distribution package is arranged for the use on Unix systems. 
	However, since the structure is rather simple, we expect that the 
	program can be compiled and executed on other platforms without 
	serious difficulties. The package is composed of the following files 
	and directories:

\begin{tabbing}
	{\tt Makefile} $\qquad$ \= : \= the {\tt Makefile} for the setup,\\
	\> \> {\tt Makefile.aix}, {\tt Makefile.linux} and 
	{\tt Makefile.solaris} are \\
	\> \> example {\tt Makefile}s for IBM-AIX, Linux and Solaris, 
	respectively,\\
	{\tt README} \> : \> a file describing how to set up the programs,\\
	{\tt VERSION-1.0} \> : \> a note for this version,\\
	{\tt 400/} - {\tt 407/} \> : \> "GRACE output codes"; the directory 
	name corresponds to \\
	\> \> the subprocess number,\\
	{\tt basesv5.1/} \> : \> BASES/SPRING (version 5.1) source codes,\\
	{\tt chanel/} \> : \> CHANEL source codes,\\
	{\tt grckinem/} \> : \> source codes of kinematics,\\
	{\tt example/} \> : \> source codes of example programs,\\
	{\tt inc/} \> : \> {\tt INCLUDE} files,\\
	{\tt lib/} \> : \> the directory to store object libraries; initially 
	empty.
\end{tabbing}

	Users have to edit the file {\tt Makefile} to specify an appropriate 
	compiler and associated compile options, as well as the paths to the 
	GR@PPA\_4b directory, and PYTHIA and CERNLIB libraries. Those parts to 
	be edited can be found at the top of the {\tt Makefile}. We prepared 
	examples for IBM-AIX, Linux and Solaris systems. All library routines 
	are compiled and combined to object libraries if users execute the 
	command {\tt make} from the GR@PPA\_4b top directory. The object 
	libraries are then moved to the directory {\tt lib/} if the command 
	{\tt make install} is executed. The {\tt Makefile}s of example 
	programs in {\tt example/} are set up by executing {\tt make example}.

\subsection{Dependencies on PYTHIA}

	GR@PPA\_4b internally uses some utility programs provided by PYTHIA. 
	The functions {\tt PYALEM} and {\tt PYALPS} are used to determine the 
	$Q^{2}$-dependent coupling strengths of QED and QCD in the matrix 
	element calculation. Since the $Q^{2}$ is given to these functions as 
	an argument, their behaviors are basically controlled by GR@PPA\_4b 
	routines. However, they require additional parameters to define the 
	running. Users are required to set relevant control parameters, such 
	as {\tt PARU(112)} before calling any initialization routines. 

	In addition, GR@PPA\_4b uses the PYTHIA function {\tt PYPDFU} for 
	referring to PYTHIA built-in PDFs. Users have to make a choice of PDF 
	by setting the parameter {\tt MSTP(51)}. The phase-space cuts defined 
	by the PYTHIA parameters {\tt CKIN(1, 2, 7, 8, 21 - 28)} are also 
	applied in GR@PPA\_4b if they are specified. These cuts are referred 
	to in the definition of the "kinematics"; namely, they limit the range 
	of kinematical variables of the final state.

\subsection{Initialization and customization}

	Although the execution of GR@PPA is controlled by the subroutine 
	{\tt grcpygen}, the detailed behavior depends on some parameters in 
	common blocks and conditions defined in some subprograms. Users can 
	change those details by changing appropriate parameters and 
	subprograms described in the following.

	The parameter that is necessary to be given by users is {\tt grcecm}, 
	which specifies the cm energy of the beam collision in GeV. 
	Optionally, users can define some phase-space cuts in the laboratory 
	frame: {\tt gptcut}, {\tt getacut} and {\tt grconcut}. These 
	parameters define the minimum $p_{T}$ in GeV, the largest 
	pseudorapidity in the absolute value and the minimum separation in 
	$\Delta R$, respectively, to be required to all produced $b$ quarks. 
	The separation ($\Delta R$) is defined for every pair of $b$ quarks as 
\begin{equation}
	\Delta R = \sqrt{\Delta\phi^{2} + \Delta\eta^{2}},
	\label{eq:DeltaR}
\end{equation}
	where $\Delta\phi$ and $\Delta\eta$ are the separation in the 
	azimuthal angle and the pseudorapidity, respectively. These parameters 
	are accessible if the file {\tt grchad.inc} in the directory 
	{\tt inc}/ is included. These cuts are applied after the kinematical 
	variables of an event are determined. In addition, users can define 
	their own cuts by editting the subroutine {\tt grcusrcut}, in which 
	4-momenta of all partons are provided through a common block. Note 
	that, since these cuts are applied during the event generation in 
	SPRING, they are smeared by later simulations in PYTHIA.

	Most of the conditions of GR@PPA\_4b are defined in the subroutine 
	{\tt grcpar}, included in the file {\tt grcpar.F} in 
	{\tt example/pyth/}. Those parameters which users are allowed to 
	change are listed in Table~\ref{grcpar}. Users can choose different 
	conditions for different subprocesses. The integer variable 
	{\tt ibswrt} controls whether BASES should be called in the 
	initialization or not. The task of BASES is to optimize the 
	integration grids and, after that, store the optimized results in a 
	"BASES table''. The execution of BASES consumes much CPU time because 
	a precise evaluation is necessary for an efficient event generation by 
	SPRING. It is not necessary to repeat the execution for identical 
	conditions. A previously optimized result ("BASES table") is reused if 
	{\tt ibswrt = 1}. It should be noted that, once the {\tt CKIN} cuts 
	and/or cuts by {\tt gptcut}, {\tt getacut}, {\tt grconcut}, and user 
	defineded-cuts are changed, the "condition" is no longer identical 
	and BASES has to be re-executed. Of course, users have to set 
	{\tt ibswrt = 0} if they change other fundamental parameters, such as 
	the cm energy, the incident beams and PDF described before, as well 
	as the energy scales and the particle masses described below.

	The variable {\tt icoup} determines the energy scale ($Q^{2}$) for 
	calculating the coupling strengths, $\alpha_{em}$ and $\alpha_{s}$, in 
	the matrix element calculation (renormalization scale). Namely, the 
	determined $Q^{2}$ is passed to {\tt PYALEM} and {\tt PYALPS}. The 
	selectable choices are listed in Table~\ref{grcpar}. The variable 
	{\tt ifact} determines $Q^{2}$ for PDF 
	(factorization scale). The definition is the same as {\tt icoup}. The 
	same choice as {\tt icoup} is taken if {\tt ifact} is not explicitly 
	given. As an option, users can apply their own definitions of these 
	energy scales, by setting {\tt icoup = 6} and/or {\tt ifact = 6} and 
	editing the subroutine {\tt grcusrsetq}. An example is attached to 
	{\tt grcpar.F}.

	The parameter {\tt ncall} specifies the number of sampling points in 
	each step of the iterative grid optimization in BASES. The larger this 
	number is, the better the conversion would be. However, it takes 
	longer in the CPU time. The optimized values are preset in 
	{\tt grcpar.F}. The character variable {\tt grcfile} gives the 
	``BASES table'' file name\footnote{BASES actually creates two files 
	having extentions of .data and .result, respectively, added to the 
	name given by {\tt grcfile}. The former is the ``BASES table'', while 
	the latter is a readable summary of the BASES execution.}. A new file 
	must be specified if {\tt ibswrt = 0}, while an existing file must be 
	specified if {\tt ibswrt = 1}. 

	The particle masses, decay widths and couplings to be used in the 
	matrix element calculation are defined in the subroutine {\tt setmas} 
	included in {\tt grcpar.F}. The mass and the total decay width of the 
	Higgs boson can be manually controlled. GR@PPA\_4b does not give any 
	constraint to these parameters. For some heavy particle masses and 
	widths, the same values are set to the corresponding PYTHIA parameters 
	in order to preserve the consistency.

	PYTHIA requires {\tt PYUPEV} users to combine the final-state partons 
	into pairs and to give the energy scale of the final-state parton 
	shower for each pair. The definition is rather trivial for those 
	subprocesses in which at least one of the $b\bar{b}$ pairs is 
	produced via a color-singlet particle production, $\gamma/Z$ and the 
	Higgs boson. On the other hand, there is not any established guiding 
	principle for pure QCD subprocesses. We give a definition based on 
	the color flow in these cases. The energy scale is set equal to the 
	invariant mass of each color-connected pair. Users can try their own 
	definitions by editing the subroutine {\tt grcxxxdetc} included in the 
	file {\tt spxxxdetc.F} in subprocess directories, where {\tt xxx} 
	denotes an {\tt ISUB} number. The energy scale of the initial-state 
	parton shower in PYHTIA is taken to be equal to the factorization 
	scale for PDF, as the default. Users can also change this definition 
	in the above subroutine if they want.

\subsection{Sample program}

	A sample program {\tt sample\_pyth.F} is provided in the subdirectory 
	{\tt example/pyth/}. Execution of the command {\tt make example} from 
	the GR@PPA\_4b top directory sets up the {\tt Makefile} for this 
	program.

	The program, first of all, sets the initialization parameters 
	described in the last section, together with some additional PYTHIA 
	parameters. After that, it calls {\tt grcpygen} for the 
	initialization. BASES optimizes the integration grids and evaluates 
	the total cross section here if {\tt ibswrt = 0}. Note that 
	{\tt grcpygen} has to be called for every subprocess that users want 
	to activate.

	The initialization of PYTHIA by {\tt PYINIT} is done after that. It is 
	necessary to set {\tt MSUB} parameters before the initialization, in 
	order to activate the subprocesses. The parameter {\tt MSEL} should 
	be set to zero.

	An event generation loop follows the initialization part. A call to 
	{\tt PYEVNT} automatically results in a call to {\tt grcpygen} through 
	{\tt PYUPEV}. The source code of {\tt PYUPEV}, dedicated to the use in 
	GR@PPA\_4b, is attached to the bottom of this sample program. The 
	generated event information is returned in the common {\tt PYJETS}. 
	This sample program outputs the information of the produced four $b$ 
	quarks as an Ntuple file. Users can obtain some histograms by 
	executing the sample macro {\tt sample.kumac} in the environment of 
	PAW. Refer to the PAW manual \cite{paw} for the usage of Ntuple and 
	PAW.

	In the output of GR@PPA\_4b, users should pay appropriate attention to 
	the print out from BASES, especially when they apply tight cuts. Since 
	each subprocess is composed of many coherent diagrams, it is not 
	practicable to take all singularities into account in the "kinematics" 
	definitions. Some very minor ones are ignored in GR@PPA\_4b. A 
	combination of very tight cuts may enhance the relative contribution 
	of ignored singularities. In such cases, it is likely to happen that, 
	in the BASES iteration, the estimated total cross section jumps 
	(increases) to a value unreasonably different from the previous 
	estimation and, accordingly, the estimated error increases. Users 
	should consider that they must be in such a trouble if they find a 
	jump of, for instance, more than three times the previous error. The 
	results are unreliable in the phase-space region defined by such cuts.
	The instructive integration accuracy is $0.5$\% or better for every 
	iteration. Users should change the parameter {\tt ncall} to a larger 
	value if this accuracy is not achieved.

\subsection{Options}

	In the default setting, GR@PPA\_4b uses one of the PYTHIA built-in 
	PDFs. We have prepared a method to refer to PDFLIB \cite{pdflib} as an 
	option. Users can switch to this option by making an appropriate 
	change in the {\tt Makefile} for building the final executable module. 
	The way how to change it is indicated in the {\tt Makefile} in the 
	subdirectory {\tt example/pyth/}. If users choose this option, the 
	coupling strength of the strong interaction for the matrix element 
	calculation is evaluated by the function {\tt ALPHAS2} in PDFLIB, 
	instead of {\tt PYALPS} of PYTHIA. In addition, a constant value is 
	used for the electroweak coupling. 

	The method to call PDFLIB by setting {\tt MSTP(52) = 2}, which is 
	described in the PYTHIA manual, is not officially supported in 
	GR@PPA\_4b, because this method requires a certain manipulation of the 
	PYTHIA library.

	In addition to the default way of using GR@PPA\_4b, where it is 
	connected to PYTHIA, we have prepared an option in which GR@PPA\_4b 
	can be executed as a stand-alone program. An example can be found in 
	the subdirectory {\tt example/alone/}. This option does not use any 
	PYTHIA subprogram. Namely, PDFLIB is used for referring to PDFs, 
	{\tt ALPHAS2} and a constant electroweak coupling are used, and 
	{\tt CKIN} cuts are not applied. All other parts are identical to the 
	default option. Therefore, one obtains an identical result, at least 
	concerning the total cross section. This option may be useful for 
	debugging.

\section{Results}

	The total cross sections estimated by GR@PPA\_4b without any cuts are 
	presented in Table {\ref{tab:xsec}} for each subprocess. The results 
	are shown for the cases of Tevatron Run-II ($p\bar{p}$ collisions at 
	$\sqrt{s}$ $=$ 2 TeV) and LHC ($pp$ collisions at $\sqrt{s}$ $=$ 14 
	TeV). We used CTEQ5L\cite{cteq} in PYTHIA 6.1 for PDF. The Higgs boson 
	mass and width are assumed to be 120 GeV/$c^{2}$ and 6.54 MeV, 
	respectively. The $b$ quark mass is set to 4.8 GeV/$c^{2}$. The 
	renormalization and factorization scales ($Q^{2}$) are chosen to be 
	identical, and those values listed in Table {\ref{tab:xsec}} are 
	assumed. The results for {\tt ISUB} = 400, 401, 405 and 406 for both 
	Tevatron Run-II and LHC conditions were found to be in good agreement 
	with corresponding results from CompHEP \cite{ilyin}.

\begin{figure}[t]
\begin{center}
\includegraphics[width=14.0cm]{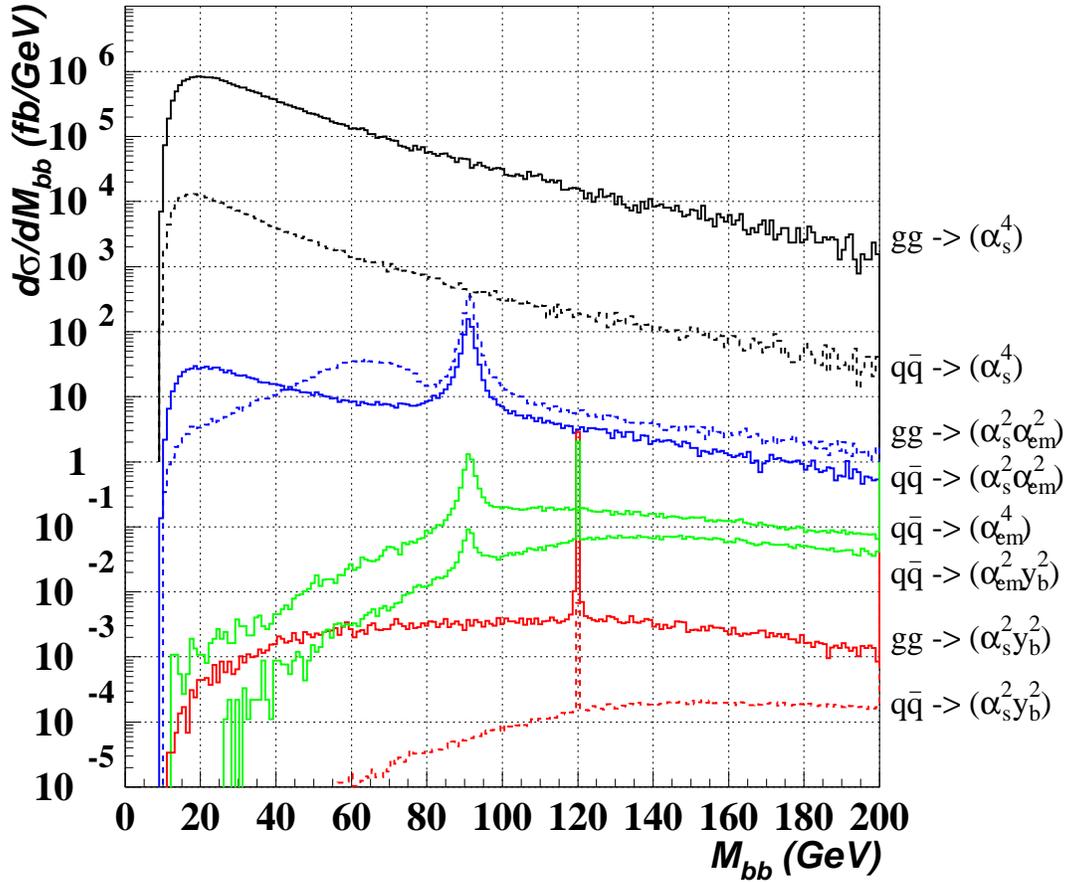}
\caption{Invariant mass distribution of leading two $b$ quarks 
        in the four-$b$ events generated by GR@PPA\_4b.
        The distribution is shown separately for each subprocess.}
\label{fig:bbmass}
\end{center}
\end{figure}

	The invariant mass distributions of two $b$ quarks having the 
	largest and the second largest transverse energy ($E_{t}$) with 
	respect to the beam direction are shown in Fig.~\ref{fig:bbmass} for 
	the Tevatron Run-II case. The results were obtained by turning off all 
	simulations by PYTHIA. The peaks corresponding to the production of 
	the Higgs boson and the $Z$ boson are clearly seen. We can also see 
	that the contribution of pure QCD subprocesses are quite large. 
	Adequate phase-space cuts and/or an appreciable enhancement are 
	necessary so that the Higgs boson signal become visible. It should be 
	noted that, in the $\alpha_{s}^{2} \alpha_{em}^{2}$ subprocesses 
	({\tt ISUB} = 402 and 403), off-resonance effects are clearly seen 
	below the $Z$-boson peak. This shows that the electroweak effects 
	(both $Z$ and $\gamma$ exchanges) are exactly evaluated in this 
	program.

	The performance of GR@PPA\_4b for the Tevatron Run-II condition is 
	summarized in Table {\ref{tab:performance}}. The integration accuracy 
	achieved in BASES is fairly better than 1\% for all subprocesses with 
	the default {\tt ncall} settings. The generation efficiency in SPRING 
	is better than a few percent for most of the subprocesses. These 
	numbers are exceptionally good for this kind of complicated processes. 
	The performance for {\tt ISUB} = 402 is apparently worse than the 
	others because the singularity structure is much complicated in this 
	subprocess. Also presented is the CPU time consumed on a Linux PC with 
	the ALPHA 700 MHz processor. The integration time and the generation 
	speed are separately shown. The generation speed does not include the 
	time consumption due to the parton shower and hadronization/decays in 
	PYTHIA. All processes except for {\tt ISUB = 402} show generation 
	speeds faster than 4 events/sec.

\section{Summary}

	We have developed an event generator, named GR@PPA\_4b, for the 
	production of four bottom quarks at $pp$ and $p\bar{p}$ collisions. 
	The program can generate events from all possible interactions at the 
	tree level within the framework of the Standard Model. The Higgs boson 
	and $\gamma$/$Z$ mediated processes as well as the pure QCD processes 
	are implemented with both $gg$ and $q\bar{q}$ ($q$ $\ne$ $b$) initial 
	states taken into acount.

	The program is based on the newly developed GR@PPA framework, an 
	extension of the GRACE automatic event-generator generation system to 
	hadron collisions. This extension allows us to incorporate the 
	variation in the initial state ($i.e.$, the partonic structure of 
	hadrons) into the GRACE system. The program includes an interface 
	to the PYTHIA 6.1 general-purpose event generator. The whole 
	GR@PPA\_4b routines can be embedded in it. This implementation allows 
	us to add simulations of the initial- and final-state parton showers, 
	hadronization and decays, as well as the additional underlying 
	activities, using PYTHIA facilities.

	This program will be useful for studies of $b\bar{b}H$ productions 
	followed by $H \rightarrow b\bar{b}$ decay. Though this process 
	will be hard to observe if $H$ is the Standard-Model Higgs boson, it 
	is expected to be greatly enhanced and become visible in some extended 
	models. The program can be used for evaluating the observability of 
	such Higgs bosons. It should be emphasized that GR@PPA\_4b can 
	generate not only the signal events, but also various irreducible 
	backgrounds. Especially, the most dangerous QCD background, which have 
	been evaluated using crude approximations so far, can be evaluated on 
	the basis of an exact tree-level calculation.

\section{Acknowledgements}

	This study was carried out in a collaboration between the Atlas-Japan 
	group formed by Japanese members of the Atlas experiment and the 
	Minami-Tateya numerical calculation group lead by Y. Shimizu. We would 
	like to thank all members of both groups. Particularly, we are 
	grateful to K. Kato for useful guidance and comments. We also thank 
	V. Ilyin for fruitful discussions.

\clearpage

\clearpage

\clearpage

\begin{table}[t]
\begin{center}
\begin{tabular}{cccc} \hline
{\tt ISUB} & Initial & Coupling & Total number \\
           &  state  &   order  &  of diagrams \\ \hline
400 & $gg$         & $\alpha_{s}^{2} y_{b}^{2}$       & 48  \\
401 & $q\bar{q}$   & $\alpha_{s}^{2} y_{b}^{2}$       & 32  \\
402 & $gg$         & $\alpha_{s}^{2} \alpha_{em}^{2}$ & 96  \\
403 & $q\bar{q}$   & $\alpha_{s}^{2} \alpha_{em}^{2}$ & 192 \\
404 & $q\bar{q}$   & $\alpha_{em}^{2} y_{b}^{2}$      & 80  \\
405 & $gg$         & $\alpha_{s}^{4}$                 & 76  \\
406 & $q\bar{q}$   & $\alpha_{s}^{4}$                 & 56  \\
407 & $q\bar{q}$   & $\alpha_{em}^{4}$                & 192 \\ \hline
\end{tabular}
\end{center}
\caption{Eight subprocesses implemented in GR@PPA\_4b are listed 
        in the order of the assigned {\tt ISUB} number.
        The subprocesses are classified according to the difference in 
        the initial-state parton combination and the coupling order.
        The total number of diagrams included are also listed for each 
        subprocess in unitary gauge.}
\label{tab:subprocess}
\end{table}


\begin{table}[htbp]
\begin{center}
\begin{tabular}{ll}
\hline
Variable      & Description \\ \hline
{\tt ibswrt}  & 0: BASES is called in the initialization. \\
              & 1: A previous BASES result is reused. \\ \hline
{\tt icoup}   & Choice of the energy scale ($Q^{2}$) for couplings \\
              & 1: $\hat{s}$ of the hard interaction \\
              & 2: Average of the squared transverse mass of $b$ quarks 
	($<m_{T}^{2}>$) \\
              & 3: Sum of the squared transverse mass of $b$ quarks 
	($\sum m_{T}^{2}$) \\
              & 4: Maximum of the squared transverse mass of $b$ quarks 
	($max$ $m_{T}^{2}$) \\
              & 5: Constant value (Set {\tt grcq} in GeV.) \\
	      & 6: User defineded energy scale defined in the subroutine 
	{\tt grcusrsetq}. \\ \hline
{\tt ifact}   & Choice of the energy scale ($Q^{2}$) for PDF \\
              & The definitions are the same as {\tt icoup}. \\
              & If {\tt ifact} = 5: Constant value (Set {\tt grcfaq} in GeV.) \\
	      & If not set explicitly, this is taken as the same as 
	{\tt icoup}. \\ \hline
{\tt ncall}   & Number of sampling points per iteration in BASES \\ \hline
{\tt grcfile} & Output file name of the BASES result \\ 
              & A new file if {\tt ibswrt} = 0; 
                       an existing file if {\tt ibswrt} = 1. \\ \hline
\end{tabular}
\end{center}
\caption{Initialization parameters to be specified in the subroutine 
        {\tt grcpar}.}
\label{grcpar}
\end{table}

\clearpage

\begin{table}[t]
\begin{center}
\begin{tabular}{cccc} \hline
{\tt ISUB} & $Q^{2}$ & $\sigma$(pb) & $\sigma$(pb) \\
 & & Tevatron II & LHC \\ \hline
400 & $M_{H}^{2}$ & 3.409(8) $\times$ 10$^{-3}$ & 5.37(1) $\times$ 10$^{-1}$ \\
401 & $M_{H}^{2}$ & 4.296(8) $\times$ 10$^{-5}$ & 4.093(5) $\times$ 10$^{-4}$ \\
402 & $M_{Z}^{2}$ & 1.924(5)                    & 1.595(4) $\times$ 10$^{2}$ \\
403 & $M_{Z}^{2}$ & 3.129(8)                    & 2.598(4) $\times$ 10$^{1}$ \\
404 & $M_{H}^{2}$+M$_{Z}^{2}$ & 1.095(2) $\times$ 10$^{-2}$ & 9.391(8) $\times$ 10$^{-2}$ \\
405 & $< m_{T}^{2} >$ & 2.588(5) $\times$ 10$^{4}$ & 7.56(1) $\times$ 10$^{5}$ \\
406 & $< m_{T}^{2} >$ & 3.413(8) $\times$ 10$^{2}$ & 1.676(1) $\times$ 10$^{3}$ \\
407 & 2$M_{Z}^{2}$ & 2.698(5) $\times$ 10$^{-2}$ & 2.568(2) $\times$ 10$^{-1}$ \\ \hline
\end{tabular}
\end{center}
\caption{The total cross section estimated by GR@PPA\_4b. 
Results are presented for the cases of Tevatron Run-II and LHC, 
with CTEQ5L in PYTHIA 6.1 for PDF and without applying any phase-space cut.
The Higgs-boson mass is assumed to be 120 GeV/$c^{2}$.}
\label{tab:xsec}
\end{table}


\begin{table}[htbp]
\begin{center}
\begin{tabular}{crcccc} \hline
{\tt ISUB} & {\tt ncall} & Integration   & Integration  & Generation 
& Generation \\
& & accuracy (\%) & time (hour) & efficiency (\%) & speed (evts/sec) \\ \hline
400 &  300000 & 0.258 &  2.31 &  2.18 &   7.5 \\
401 &   10000 & 0.204 &  0.01 & 53.39 &   834 \\
402 & 1000000 & 0.274 & 27.38 &  0.52 &  0.78 \\
403 &  200000 & 0.280 &  4.00 &  3.30 &   4.4 \\
404 &   20000 & 0.203 &  0.03 & 37.94 &   386 \\
405 &  100000 & 0.207 &  2.58 &  6.76 &   6.6 \\
406 &   10000 & 0.238 &  1.57 & 48.05 &   265 \\
407 &  100000 & 0.206 &  0.97 & 10.67 &    27 \\ \hline
\end{tabular}
\end{center}
\caption{The performance of GR@PPA\_4b is summarized for each subprocess. 
	   Tests were done using a Linux PC (ALPHA 700 MHz) in the Tevatron 
	   Run-II condition, with CTEQ5L in PYTHIA 6.1 for PDF and without 
	   applying any phase-space cut. The integration accuracy and the 
	   integration time are relevant to the execution of BASES at the 
	   initialization stage. The generation efficiency and the generation 
	   speed show the performance of the SPRING execution. The generation 
	   speed does not include the time consumed for the parton showers, 
	   hadronization and decays by PYTHIA.}
\label{tab:performance}
\end{table}

\clearpage

\appendix
\section{Feynman diagrams}

\begin{figure}[htbp]
\begin{center}
\includegraphics[width=7.0cm]{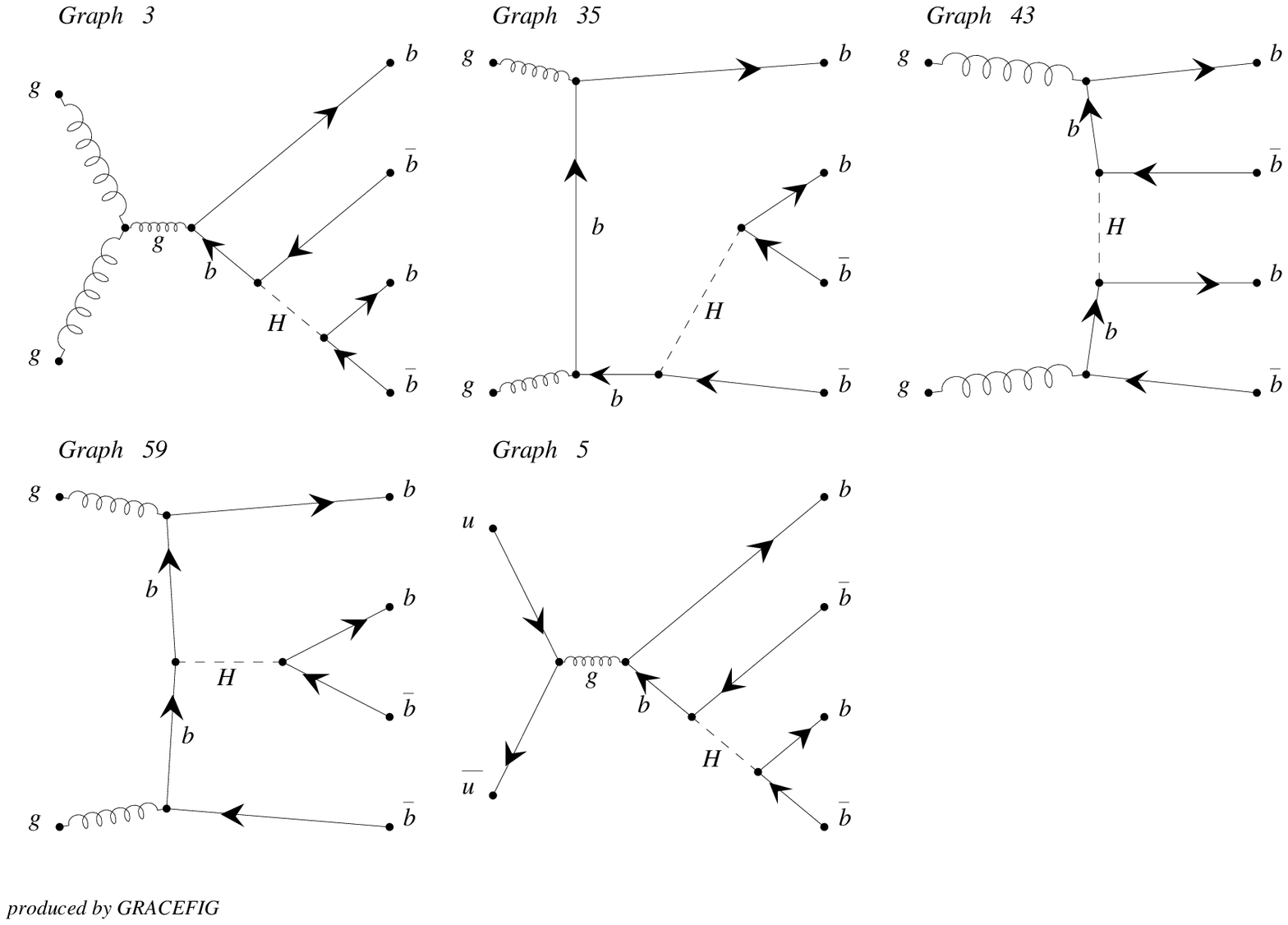}
\caption{Typical Feynman diagrams having a coupling order of 
        $\alpha_{s}^{2}y_{b}^{2}$.}
\label{bbbb_h2q2_fyn}
\end{center}
\end{figure}

\begin{figure}[htbp]
\begin{center}
\includegraphics[width=7.0cm]{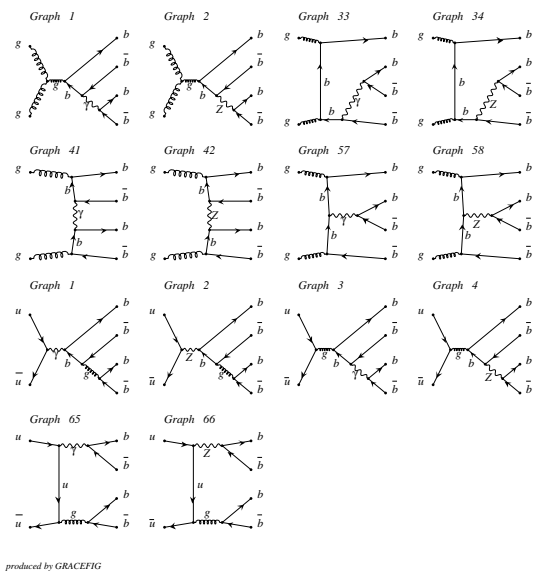}
\caption{Typical Feynman diagrams having a coupling order of
        $\alpha_{s}^{2}\alpha_{em}^{2}$.}
\label{bbbb_e2q2_fyn}
\end{center}
\end{figure}

\begin{figure}[htbp]
\begin{center}
\includegraphics[width=8.0cm]{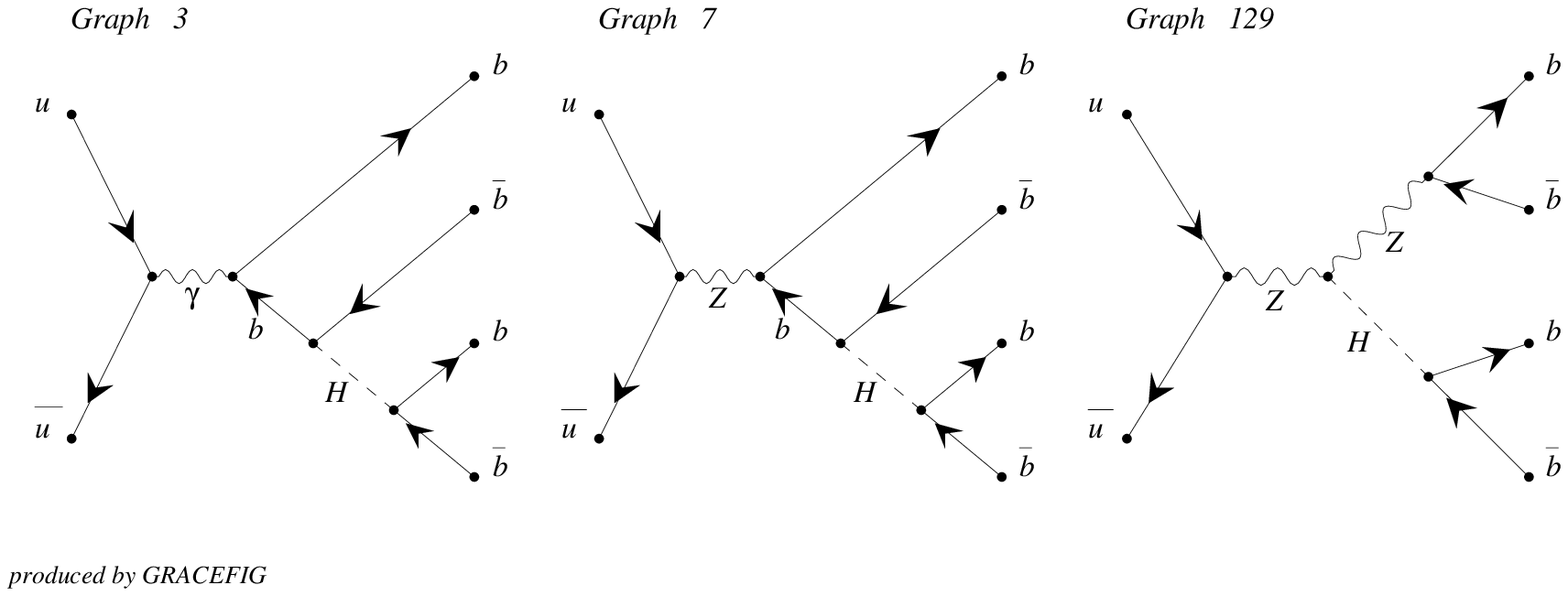}
\caption{Typical Feynman diagrams having a coupling order of
        $\alpha_{em}^{2}y_{b}^{2}$.}
\label{bbbb_h2e2_fyn}
\end{center}
\end{figure}

\begin{figure}[htbp]
\begin{center}
\includegraphics[width=8.0cm]{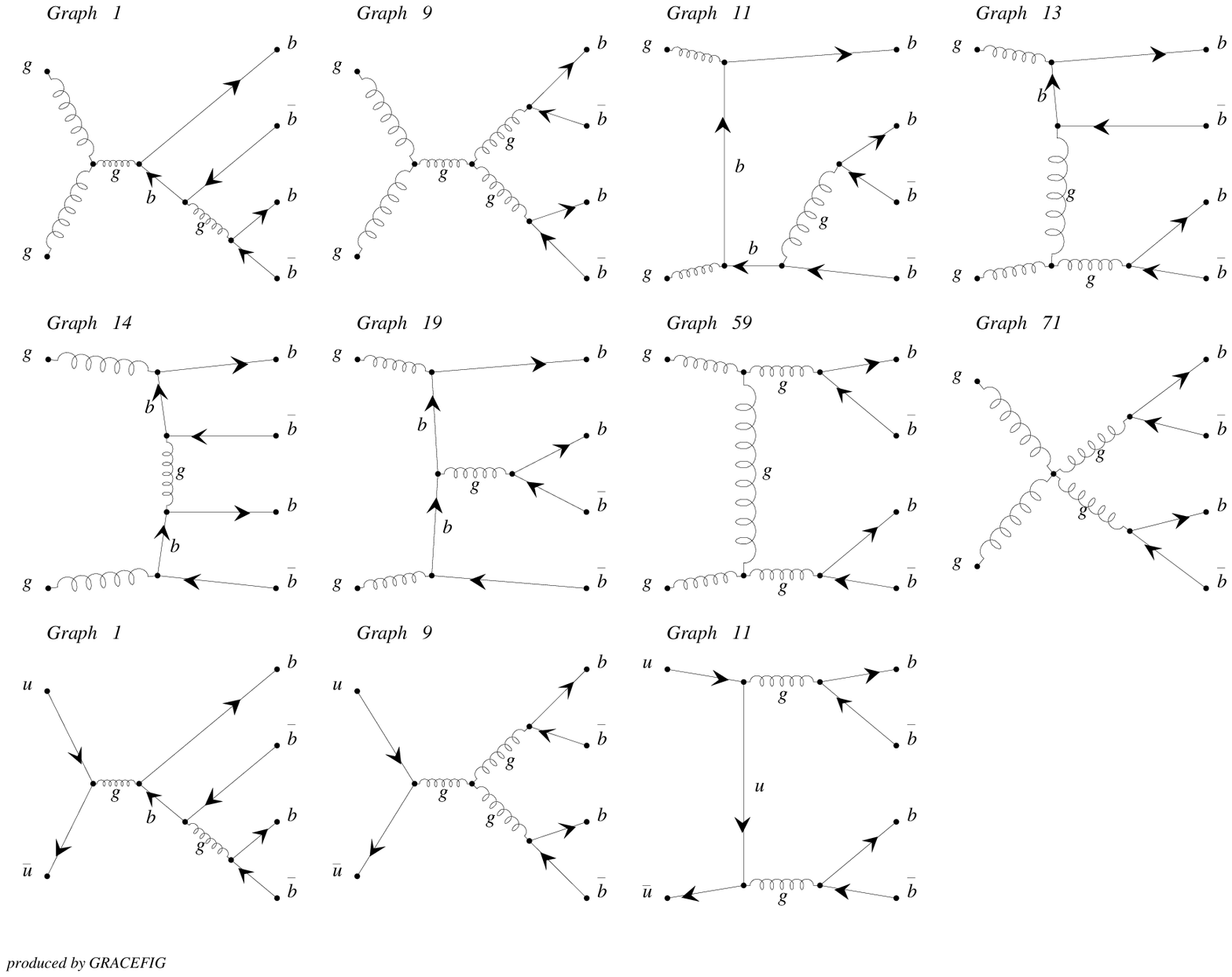}
\caption{Typical Feynman diagrams having a coupling order of 
        $\alpha_{s}^{4}$.}
\label{bbbb_q4_fyn}
\end{center}
\end{figure}

\begin{figure}[htbp]
\begin{center}
\includegraphics[width=8.0cm]{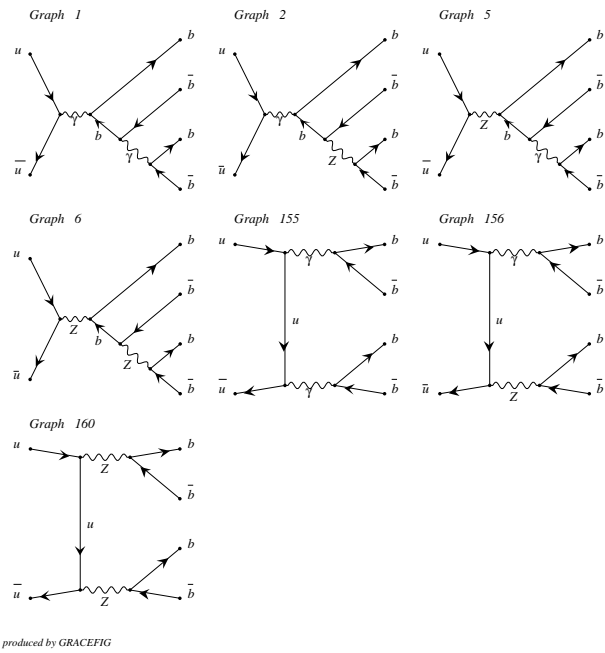}
\caption{Typical Feynman diagrams having a coupling order of 
        $\alpha_{em}^{4}$.}
\label{bbbb_e4_fyn}
\end{center}
\end{figure}

        Typical Feynman diagrams of the interactions implemented in GR@PPA\_4b 
        are illustrated in Figs.~\ref{bbbb_h2q2_fyn}-\ref{bbbb_e4_fyn}.

\clearpage


\clearpage

\section{Sample code}

{\small
\renewcommand{\baselinestretch}{0.78}
\begin{verbatim}
C Program : sample_pyth.F
C Purpose : Sample code to connect to PYTHIA 6.1xx
C Date    : Dec.10.2001
C Author  : Soushi Tsuno

C Only those subprocesses in which a Higgs boson decaying to a 
C b-quark pair is produced in association with a b-quark pair 
C production via QCD (ISUB = 400 and 401) are activated in this 
C example.

C...All real arithmetic in double precision.
      implicit double precision(a-h,o-z)
C...Three Pythia functions return integers, so need declaring.
      integer pyk,pychge,pycomp
C...EXTERNAL statement links PYDATA on most machines.
      external pydata
C...Commonblocks.
      common/pyjets/n,npad,k(4000,5),p(4000,5),v(4000,5)
      common/pypars/mstp(200),parp(200),msti(200),pari(200)
      common/pydat1/mstu(200),paru(200),mstj(200),parj(200)
      common/pysubs/msel,mselpd,msub(500),kfin(2,-40:40),ckin(200)

C...Counter for number of generated events of each type.
      dimension ncount(8)
      data ncount/8*0/

C...Include GR@PPA common parameter.
      include './inc/grchad.inc'

C...Set Ntuple-----------------------------------------------------
      integer nwpawc
      parameter (nwpawc=300000)
      common/pawc/paw(nwpawc)
      integer neve,psub
      real pj
      common/genep/neve,psub,pj(5,6)

      call hlimit(nwpawc)
      call hropen(1,'genep','bbbb_tev.nt','N',4095,istat)
      call hbnt(10,'genep',' ')
      call hbname(10,'genep',neve,'neve:i,psub:i,pj(5,6):r')
C-----------------------------------------------------------------

C...Number of events and cm energy. 
      nev = 1000                ! Number of events
      grcecm = 2000.0d0         ! CM Energy

C...Kinematical cuts.
      gptcut   = 0.0d0          ! Pt Cut for each particles
      getacut  = 100.0d0        ! Eta Cut
      grconcut = 0.0d0          ! RCone Cut
C   Other cuts can be applies using CKIN parameters of Pythia.
C   Furthermore, users can define their own cuts by editing the 
C   subroutine grcusrcut in grcpar.F.

C...PDF and Coupling.
C   These parameters have to be set before the GR@PPA_4b initialization.
      mstp(51)  = 7             ! CTEC5L
      mstp(58)  = 4             ! Nr. of flavor in PDF
      mstu(101) = 1             ! running alpha_em
      iord_als  = 1             ! first-order running alpha_s
      nfrv_als  = 5             ! Nr. of flavors assumed in alpha_s
      aLam_als  = 0.146d0       ! Lambda used in alpha_s

      mstu(111) = iord_als
      mstu(112) = nfrv_als
      paru(112) = aLam_als

C...GR@PPA_4b initialization (BASES integration).
      call grcpygen('PAP',400,1,sigmax) ! gg -> h0(bb)+bb(Higgs2,QCD2)
      call grcpygen('PAP',401,1,sigmax) ! qq -> h0(bb)+bb(Higgs2,QCD2)

C...Pythia initialization.
      msel = 0
      msub(400) = 1         ! External process on
      msub(401) = 1         ! External process on
C   Switch off unnecessary aspects of Pythia.
C      mstp(61)  = 0             ! Initial state radiation OFF
C      mstp(71)  = 0             ! Final state radiation OFF
C      mstp(81)  = 0             ! Multiple interaction OFF
C      mstp(111) = 0             ! Hadronization OFF
C   Initialization.
      call pyinit('CMS','p','pbar',grcecm)

C...The alpha_s parameters must be set again here if they are non-
C   standard. These parameters are over-written with standard ones 
C   in pyinit according the choice of PDF.
      mstu(111) = iord_als
      mstu(112) = nfrv_als
      paru(112) = aLam_als

C...Event loop.
      do iev = 1,nev

         call pyevnt

         isub = msti(1)
         icase = 1
         if (isub.ge.400 .and. isub.le.407) icase = isub - 399
         ncount(icase) = ncount(icase) + 1
         if (ncount(icase).le.1) then
            write(6,*) ' Following event is subprocess',isub
            call pylist(1)
         endif

         if (mod(iev,1000).eq.0) then
            write(*,*) iev
         endif

C...Data Store.

C...Event Nr..
         neve = iev
C...Process ID.
         psub = isub

C...Set Particle 1,2 (x,y,z).
         do i = 1,2
            do j = 1,4
               pj(j,i) = p(i+2,j)
            enddo
            pj(5,i) = k(i+2,2)
         enddo

C...Set Particle 3,4,5,6 (pt,phi,eta).
         do i = 3,6
            pj(1,i) = pyp(i+4,10)
            pj(2,i) = pyp(i+4,15)
            pj(3,i) = pyp(i+4,19)
            pj(4,i) = pyp(i+4,4)
            pj(5,i) = k(i+4,2)
         enddo

C...Fill ntuple.
         call hfnt(10)

C...End of loop over events.
      enddo

C...Cross section table.
      call pystat(1)

C...Close ntuple.
      call hrout(10,genep,' ')
      call hrend('genep')

      end


C###########################################################
      subroutine pyupev(isub,sigev)
      implicit double precision(a-h,o-z)
      integer isub
      real*8 sigev

C...Only call GR@PPA_4b.
      call grcpygen(' ',isub,0,sigev)

      return
      end
\end{verbatim}
}

\clearpage

\section{Test run output}

{\small
\renewcommand{\baselinestretch}{0.78}
\begin{verbatim}
 ******************************************************************************
 ******************************************************************************
 **  Welcome to                                                              **
 **                                                                          **
 **       GGG  RRRR          PPPP  PPPP   AAA                                **
 **      G   G R   R   @@@@  P   P P   P A   A                               **
 **      G     R   R  @ @@ @ P   P P   P A   A                               **
 **      G GGG RRRR  @ @ @ @ PPPP  PPPP  AAAAA                               **
 **      G   G R  R  @ @@ @  P     P     A   A                               **
 **       GGG  R   R  @@@@@@ P     P     A   A  _4b                          **
 **     =======================================                              **
 **       GRace At Proton-Proton/Anti-proton                                 **
 **                                                                          **
 **       This is GR@PPA version 1.0                                         **
 **       coded by S.Tsuno (tsuno@fnal.gov)                                  **
 **       with Minami Tateya Collab. and ATLAS-J.                            **
 **                                                                          **
 **       On web...  http://www.kek.jp/                                      **
 **       Referances, ........                                               **
 **                                                                          **
 ******************************************************************************
 ******************************************************************************
 Accepted CM Energy :   2000.00000000000      GeV
          Beam type : Proton-Anti-Proton Collision
 Process : [         400 ] gg -> h0(bb)+bb(Higgs2,QCD2)
 Set BASES file
    Filename : bases_400_mh120.result
               bases_400_mh120.data 
           4  body final state
   Pt Cut of Particle           1  : none
   Pt Cut of Particle           2  : none
   Pt Cut of Particle           3  : none
   Pt Cut of Particle           4  : none
   Eta Cut of Particle           1  : none
   Eta Cut of Particle           2  : none
   Eta Cut of Particle           3  : none
   Eta Cut of Particle           4  : none
   Rcone Cut of Particle           1  : none
   Rcone Cut of Particle           2  : none
   Rcone Cut of Particle           3  : none
   Rcone Cut of Particle           4  : none
   Missing Pt(Et) Cut : none
   Option of Renormalization Scale :           5   120.000000000000      GeV
   Option of   Factorization Scale :           0
 
 Start BASES Integration!!


                                                       Date:  2/ 2/20  05:03
        **********************************************************
        *                                                        *
        *     BBBBBBB     AAAA     SSSSSS   EEEEEE   SSSSSS      *
        *     BB    BB   AA  AA   SS    SS  EE      SS    SS     *
        *     BB    BB  AA    AA  SS        EE      SS           *
        *     BBBBBBB   AAAAAAAA   SSSSSS   EEEEEE   SSSSSS      *
        *     BB    BB  AA    AA        SS  EE            SS     *
        *     BB    BB  AA    AA  SS    SS  EE      SS    SS     *
        *     BBBB BB   AA    AA   SSSSSS   EEEEEE   SSSSSS      *
        *                                                        *
        *                   BASES Version 5.1                    *
        *           coded by S.Kawabata KEK, March 1994          *
        **********************************************************

     <<   Parameters for BASES    >>

      (1) Dimensions of integration etc.
          # of dimensions :    Ndim    =       10   ( 50 at max.)
          # of Wilds      :    Nwild   =       10   ( 15 at max.)
          # of sample points : Ncall   =   299008(real)   300000(given)
          # of subregions    : Ng      =       50 / variable
          # of regions       : Nregion =        2 / variable
          # of Hypercubes    : Ncube   =     1024

      (2) About the integration variables
          ------+---------------+---------------+-------+-------
              i       XL(i)           XU(i)       IG(i)   Wild
          ------+---------------+---------------+-------+-------
              1    0.000000E+00    1.000000E+00     1      yes
              2    0.000000E+00    1.000000E+00     1      yes
              3    0.000000E+00    1.000000E+00     1      yes
              4    0.000000E+00    1.000000E+00     1      yes
              5    0.000000E+00    1.000000E+00     1      yes
              6    0.000000E+00    1.000000E+00     1      yes
              7    0.000000E+00    1.000000E+00     1      yes
              8    0.000000E+00    1.000000E+00     1      yes
              9    0.000000E+00    1.000000E+00     1      yes
             10    0.000000E+00    1.000000E+00     1      yes
          ------+---------------+---------------+-------+-------

      (3) Parameters for the grid optimization step
          Max.# of iterations: ITMX1 =        5
          Expected accuracy  : Acc1  =   0.2000 %

      (4) Parameters for the integration step
          Max.# of iterations: ITMX2 =        5
          Expected accuracy  : Acc2  =   0.0100 %


                                                       Date:  2/ 2/20  05:03
               Convergency Behavior for the Grid Optimization Step
 ------------------------------------------------------------------------------
 <- Result of  each iteration ->  <-     Cumulative Result     -> < CPU  time >
  IT Eff R_Neg   Estimate  Acc %  Estimate(+- Error )order  Acc % ( H: M: Sec )
 ------------------------------------------------------------------------------
   1  99  0.00  3.417E-03  0.902  3.417196(+-0.030830)E-03  0.902   0:14: 2.08
   2  99  0.00  3.460E-03  0.669  3.444653(+-0.018513)E-03  0.537   0:28: 4.14
   3  99  0.00  3.421E-03  0.583  3.433936(+-0.013571)E-03  0.395   0:42: 6.15
   4  99  0.00  3.404E-03  0.573  3.424234(+-0.011139)E-03  0.325   0:56: 8.20
   5  99  0.00  3.438E-03  0.572  3.427558(+-0.009692)E-03  0.283   1:10:10.26
 ------------------------------------------------------------------------------


                                                       Date:  2/ 2/20  05:03
               Convergency Behavior for the Integration Step      
 ------------------------------------------------------------------------------
 <- Result of  each iteration ->  <-     Cumulative Result     -> < CPU  time >
  IT Eff R_Neg   Estimate  Acc %  Estimate(+- Error )order  Acc % ( H: M: Sec )
 ------------------------------------------------------------------------------
   1  99  0.00  3.407E-03  0.579  3.406833(+-0.019711)E-03  0.579   1:24:12.01
   2  99  0.00  3.378E-03  0.580  3.392381(+-0.013901)E-03  0.410   1:38:13.74
   3  99  0.00  3.424E-03  0.578  3.402792(+-0.011373)E-03  0.334   1:52:15.38
   4  99  0.00  3.423E-03  0.574  3.407759(+-0.009841)E-03  0.289   2: 6:16.63
   5  99  0.00  3.416E-03  0.570  3.409380(+-0.008783)E-03  0.258   2:20:18.34
 ------------------------------------------------------------------------------


                    ****** END OF BASES *********

     <<   Computing Time Information   >>

               (1) For BASES              H: M:  Sec
                   Overhead           :   0: 0: 0.00
                   Grid Optim. Step   :   1:10:10.26
                   Integration Step   :   1:10: 8.08
                   Go time for all    :   2:20:18.34

               (2) Expected event generation time
                   Expected time for 1000 events :      4.02 Sec
 ******************************************************************************
 ******************************************************************************
 **  Welcome to                                                              **
 **                                                                          **
 **       GGG  RRRR          PPPP  PPPP   AAA                                **
 **      G   G R   R   @@@@  P   P P   P A   A                               **
 **      G     R   R  @ @@ @ P   P P   P A   A                               **
 **      G GGG RRRR  @ @ @ @ PPPP  PPPP  AAAAA                               **
 **      G   G R  R  @ @@ @  P     P     A   A                               **
 **       GGG  R   R  @@@@@@ P     P     A   A  _4b                          **
 **     =======================================                              **
 **       GRace At Proton-Proton/Anti-proton                                 **
 **                                                                          **
 **       This is GR@PPA version 1.0                                         **
 **       coded by S.Tsuno (tsuno@fnal.gov)                                  **
 **       with Minami Tateya Collab. and ATLAS-J.                            **
 **                                                                          **
 **       On web...  http://www.kek.jp/                                      **
 **       Referances, ........                                               **
 **                                                                          **
 ******************************************************************************
 ******************************************************************************
 Accepted CM Energy :   2000.00000000000      GeV
          Beam type : Proton-Anti-Proton Collision
 Process : [         401 ] qq -> h0(bb)+bb(Higgs2,QCD2)
 Set BASES file
    Filename : bases_401_mh120.result                    
               bases_401_mh120.data  
           4  body final state
   Pt Cut of Particle           1  : none
   Pt Cut of Particle           2  : none
   Pt Cut of Particle           3  : none
   Pt Cut of Particle           4  : none
   Eta Cut of Particle           1  : none
   Eta Cut of Particle           2  : none
   Eta Cut of Particle           3  : none
   Eta Cut of Particle           4  : none
   Rcone Cut of Particle           1  : none
   Rcone Cut of Particle           2  : none
   Rcone Cut of Particle           3  : none
   Rcone Cut of Particle           4  : none
   Missing Pt(Et) Cut : none
   Option of Renormalization Scale :           5   120.000000000000      GeV
   Option of   Factorization Scale :           0
 
 Start BASES Integration!!


                                                       Date:  2/ 2/20  07:24
        **********************************************************
        *                                                        *
        *     BBBBBBB     AAAA     SSSSSS   EEEEEE   SSSSSS      *
        *     BB    BB   AA  AA   SS    SS  EE      SS    SS     *
        *     BB    BB  AA    AA  SS        EE      SS           *
        *     BBBBBBB   AAAAAAAA   SSSSSS   EEEEEE   SSSSSS      *
        *     BB    BB  AA    AA        SS  EE            SS     *
        *     BB    BB  AA    AA  SS    SS  EE      SS    SS     *
        *     BBBB BB   AA    AA   SSSSSS   EEEEEE   SSSSSS      *
        *                                                        *
        *                   BASES Version 5.1                    *
        *           coded by S.Kawabata KEK, March 1994          *
        **********************************************************

     <<   Parameters for BASES    >>

      (1) Dimensions of integration etc.
          # of dimensions :    Ndim    =       11   ( 50 at max.)
          # of Wilds      :    Nwild   =       11   ( 15 at max.)
          # of sample points : Ncall   =     8192(real)    10000(given)
          # of subregions    : Ng      =       50 / variable
          # of regions       : Nregion =        2 / variable
          # of Hypercubes    : Ncube   =     2048

      (2) About the integration variables
          ------+---------------+---------------+-------+-------
              i       XL(i)           XU(i)       IG(i)   Wild
          ------+---------------+---------------+-------+-------
              1    0.000000E+00    1.000000E+00     1      yes
              2    0.000000E+00    1.000000E+00     1      yes
              3    0.000000E+00    1.000000E+00     1      yes
              4    0.000000E+00    1.000000E+00     1      yes
              5    0.000000E+00    1.000000E+00     1      yes
              6    0.000000E+00    1.000000E+00     1      yes
              7    0.000000E+00    1.000000E+00     1      yes
              8    0.000000E+00    1.000000E+00     1      yes
              9    0.000000E+00    1.000000E+00     1      yes
             10    0.000000E+00    1.000000E+00     1      yes
             11    0.000000E+00    1.000000E+00     1      yes
          ------+---------------+---------------+-------+-------

      (3) Parameters for the grid optimization step
          Max.# of iterations: ITMX1 =        5
          Expected accuracy  : Acc1  =   0.2000 %

      (4) Parameters for the integration step
          Max.# of iterations: ITMX2 =        5
          Expected accuracy  : Acc2  =   0.0100 %


                                                       Date:  2/ 2/20  07:24
               Convergency Behavior for the Grid Optimization Step
 ------------------------------------------------------------------------------
 <- Result of  each iteration ->  <-     Cumulative Result     -> < CPU  time >
 ------------------------------------------------------------------------------
   1  99  0.00  4.288E-05  1.449  4.287773(+-0.062123)E-05  1.449   0: 0: 2.62
   2  99  0.00  4.291E-05  0.620  4.290673(+-0.024461)E-05  0.570   0: 0: 5.25
   3  99  0.00  4.311E-05  0.457  4.303146(+-0.015345)E-05  0.357   0: 0: 7.88
   4  99  0.00  4.311E-05  0.455  4.306190(+-0.012091)E-05  0.281   0: 0:10.50
   5  99  0.00  4.290E-05  0.430  4.301289(+-0.010110)E-05  0.235   0: 0:13.12
 ------------------------------------------------------------------------------


                                                       Date:  2/ 2/20  07:24
               Convergency Behavior for the Integration Step      
 ------------------------------------------------------------------------------
 <- Result of  each iteration ->  <-     Cumulative Result     -> < CPU  time >
  IT Eff R_Neg   Estimate  Acc %  Estimate(+- Error )order  Acc % ( H: M: Sec )
 ------------------------------------------------------------------------------
   1  99  0.00  4.296E-05  0.446  4.295959(+-0.019145)E-05  0.446   0: 0:15.74
   2  99  0.00  4.295E-05  0.458  4.295262(+-0.013719)E-05  0.319   0: 0:18.36
   3  99  0.00  4.316E-05  0.482  4.301655(+-0.011453)E-05  0.266   0: 0:20.98
   4  99  0.00  4.291E-05  0.452  4.298948(+-0.009860)E-05  0.229   0: 0:23.60
   5  99  0.00  4.290E-05  0.430  4.297026(+-0.008694)E-05  0.202   0: 0:26.22
 ------------------------------------------------------------------------------


                    ****** END OF BASES *********

     <<   Computing Time Information   >>

               (1) For BASES              H: M:  Sec
                   Overhead           :   0: 0: 0.00
                   Grid Optim. Step   :   0: 0:13.12
                   Integration Step   :   0: 0:13.09
                   Go time for all    :   0: 0:26.22

               (2) Expected event generation time
                   Expected time for 1000 events :      0.46 Sec
1                                                                              
 ******************************************************************************
 ******************************************************************************
 **                                                                          **
 **                                                                          **
 **              *......*                  Welcome to the Lund Monte Carlo!  **
 **         *:::!!:::::::::::*                                               **
 **      *::::::!!::::::::::::::*          PPP  Y   Y TTTTT H   H III   A    **
 **    *::::::::!!::::::::::::::::*        P  P  Y Y    T   H   H  I   A A   **
 **   *:::::::::!!:::::::::::::::::*       PPP    Y     T   HHHHH  I  AAAAA  **
 **   *:::::::::!!:::::::::::::::::*       P      Y     T   H   H  I  A   A  **
 **    *::::::::!!::::::::::::::::*!       P      Y     T   H   H III A   A  **
 **      *::::::!!::::::::::::::* !!                                         **
 **      !! *:::!!:::::::::::*    !!       This is PYTHIA version 6.138      **
 **      !!     !* -><- *         !!       Last date of change:  2 Mar 2000  **
 **      !!     !!                !!                                         **
 **      !!     !!                !!       Now is 20 Feb 2002 at  7:24:31    **
 **      !!                       !!                                         **
 **      !!        ep             !!       Disclaimer: this program comes    **
 **      !!                       !!       without any guarantees. Beware    **
 **      !!                 pp    !!       of errors and use common sense    **
 **      !!   e+e-                !!       when interpreting results.        **
 **      !!                       !!                                         **
 **      !!                                Copyright T. Sjostrand (1999)     **
 **                                                                          **
 ** An archive of program versions and documentation is found on the web:    **
 ** http://www.thep.lu.se/~torbjorn/Pythia.html                              **
 **                                                                          **
 ** When you cite this program, currently the official reference is          **
 ** T. Sjostrand, Computer Physics Commun. 82 (1994) 74.                     **
 ** The supersymmetry extensions are described in                            **
 ** S. Mrenna, Computer Physics Commun. 101 (1997) 232                       **
 ** Also remember that the program, to a large extent, represents original   **
 ** physics research. Other publications of special relevance to your        **
 ** studies may therefore deserve separate mention.                          **
 **                                                                          **
 ** Main author: Torbjorn Sjostrand; Department of Theoretical Physics 2,    **
 **   Lund University, Solvegatan 14A, S-223 62 Lund, Sweden;                **
 **   phone: + 46 - 46 - 222 48 16; e-mail: torbjorn@thep.lu.se              **
 ** SUSY author: Stephen Mrenna, Physics Department, UC Davis,               **
 **   One Shields Avenue, Davis, CA 95616, USA;                              **
 **   phone: + 1 - 530 - 752 - 2661; e-mail: mrenna@physics.ucdavis.edu      **
 **                                                                          **
 **                                                                          **
 ******************************************************************************
 ******************************************************************************
1****************** PYINIT: initialization of PYTHIA routines *****************

 ==============================================================================
 I                                                                            I
 I             PYTHIA will be initialized for a p on pbar collider            I
 I                  at   2000.000 GeV center-of-mass energy                   I
 I                                                                            I
 ==============================================================================

 ******** PYMAXI: summary of differential cross-section maximum search ********

           ==========================================================
           I                                      I                 I
           I  ISUB  Subprocess name               I  Maximum value  I
           I                                      I                 I
           ==========================================================
           I                                      I                 I
           I   96   Semihard QCD 2 -> 2           I    6.6438D+02   I
           I  400   gg -> h0(bb)+bb(Higgs2,QCD2)  I    3.4094D-12   I
           I  401   qq -> h0(bb)+bb(Higgs2,QCD2)  I    4.2971D-14   I
           I                                      I                 I
           ==========================================================

 ********************** PYINIT: initialization completed **********************
  Following event is subprocess         400


  !!! Event lists by PYTHIA are omitted....


1********* PYSTAT:  Statistics on Number of Events and Cross-sections *********

 ==============================================================================
 I                                  I                            I            I
 I            Subprocess            I      Number of points      I    Sigma   I
 I                                  I                            I            I
 I----------------------------------I----------------------------I    (mb)    I
 I                                  I                            I            I
 I N:o Type                         I    Generated         Tried I            I
 I                                  I                            I            I
 ==============================================================================
 I                                  I                            I            I
 I   0 All included subprocesses    I         1000          1000 I  3.452D-12 I
 I 400 gg -> h0(bb)+bb(Higgs2,QCD2) I          990           990 I  3.409D-12 I
 I 401 qq -> h0(bb)+bb(Higgs2,QCD2) I           10            10 I  4.297D-14 I
 I                                  I                            I            I
 ==============================================================================

 ********* Fraction of events that fail fragmentation cuts =  0.00000 *********

\end{verbatim}
}

\end{document}